\documentclass[
aps,
pra,
letterpaper,
accepted=2026-09-29
]{quantumarticle}
\pdfoutput=1

\usepackage{hyperref}
\usepackage{float}
\usepackage{graphicx}
\usepackage{tikz}
\usepackage{pgfplots}
\usepackage{amsmath}
\usepackage{amsfonts}
\usepackage{amssymb}
\usepackage{subcaption}
\usepackage{dcolumn}
\usepackage{bm}
\usetikzlibrary{backgrounds,fit,decorations.pathreplacing,patterns}

\newcommand{\ket}[1]{\ensuremath{\left| #1\right\rangle}}
\pgfdeclarelayer{background}
\pgfdeclarelayer{foreground}
\pgfsetlayers{background,main,foreground}

\pgfplotsset{compat=1.18}
\begin{document}

\title{Rare Event Simulation of Quantum Error-Correcting Circuits}

\author{Carolyn Mayer}
 \email{cdmayer@sandia.gov}
 \affiliation{
    Quantum Computer Science 1,
    Sandia National Laboratories, Albuquerque, NM, USA 
}
\author{Anand Ganti}
 \email{aganti@sandia.gov}
 \affiliation{
    Cyber Security Initiatives,
    Sandia National Laboratories, Albuquerque, NM, USA
}

\author{Uzoma Onunkwo}
 \email{uonunkw@sandia.gov}
 \affiliation{
    Cyber Security Initiatives,
    Sandia National Laboratories, Albuquerque, NM, USA
}
\author{Tzvetan Metodi}
\email{tsmetod@sandia.gov}
\affiliation{
Quantum Computer Science, Sandia National Laboratories,
Albuquerque, NM, USA
}

\author{Benjamin Anker}
\email{banker@unm.edu}
\affiliation{
Center for Quantum Information and Control,
University of New Mexico, Albuquerque, NM, USA
}
\affiliation{
Electrical and Computer Engineering,
University of New Mexico, Albuquerque, NM, USA
}
\author{Jacek Skryzalin}
\affiliation{
Jump Trading Group,
1 London Wall, Barbican,
London, UK
}


\begin{abstract}
We describe a practical approach for accessing the logical failure rates of quantum error-correcting (QEC) circuits under low physical (component) failure rate regimes. Standard Monte Carlo is often the de facto approach for studying the failure rates of quantum circuits. However, in the study of fault-tolerant error-correcting circuits, the ability to extend this approach to low physical failure rates is limited. In particular, the use of Monte Carlo to access circuits that are relatively large or have high correcting power becomes more difficult as we lower the input failure rates of the individual components (gates) in the circuit. For these reasons, many simulations studying the circuit model go no lower than end-to-end logical failure rates in the $10^{-6}$ regime.

In this report, we outline an approach that borrows from earlier work by Bravyi and Vargo \cite{bv13} to the more complex circuit noise model. Earlier works studied both the capacity and phenomenological noise models \cite{bv13,Beverland2019,bonilla2021xzzx}, but the work is insufficient for generating similar simulations in the circuit-noise model. To the best of our knowledge, our team is the first to develop a full prescription of the rare event simulation by splitting technique for the circuit-based noise model. We have also generated promising results that are confirmed by standard Monte Carlo simulation under an accessible regime. This work shows that we can access noise in the circuit-model prescription of quantum error-correcting code to failure rates below $10^{-20}$ regime.
\end{abstract}
\maketitle

\section{\label{sec:intro}Introduction}
In recent years,  strong interest has grown in the area of quantum computing.  This has been made
apparent by the amount of research funding and commercial sectors focusing on advancing the science for
realizing quantum computing systems.  These interests are  motivated by opportunities to better
arrive at solutions to  challenging technological problems in the areas of pharmaceutical research, cybersecurity,
and quantum chemistry to name a few.

However, the advancement towards quantum computing is deemed by most to be unattainable without quantum
error correction and fault-tolerant systems.  Many quantum computing paradigms rely on qubits, the basis
of quantum information, which are quite unreliable.  Current state-of-the-art systems are
demonstrating systems with hundreds to thousands of qubits, capable of performing thousands of primitive quantum gate operations. \cite{google2025quantum, abughanem2025ibm, di2024quantum, manetsch2025tweezer}.
However, the application of quantum
gates introduces sources of decoherence for these qubits, which must be protected.  Quantum error correction (QEC)
provides an avenue to protect qubits against environmental noise.  With many of the advances over
the last decade in hardware designs for quantum systems, today's systems are still only pushing towards $10^{-3}$
infidelity for quantum operations.  However, many experts believe that infidelity rates in the $10^{-12}$,
so-called \emph{teraquop}, range are needed for quantum systems to show real advantage over today's dominant
classical computers.  QEC, along with fault-tolerant designs, could bring such hardware designs to enable quantum
computing in the teraquop range.  Popular QEC codes, such as surface
codes~\cite{dennis2002topological,google2023suppressing,PhysRevA.86.032324,horsman2012surface,varbanov2025neural,gidney2025yoked,delfosse2014decoding} and
quantum LDPC codes~\cite{breuckmann2021quantum,panteleev2021quantum,dinur2023good,delfosse2022toward},  show promise.

Simulation software has become imperative for assessing the performance of QEC designs on different codes and their
associated decoders.  These simulations enable researchers to understand the logical failure rates under different
noise characteristics.  Many works show simulation results under varying noise constraints that range from the code
capacity noise model to the phenomenological noise model and the more realistic circuit noise model.  The assessment of
QEC designs usually leverages Monte Carlo style of simulations, but na\"{i}ve approaches can limit the ability to
assess codes in the teraquop regime.  In this work, we illustrate how we can assess codes using a method pioneered
by Bravyi and Vargo (\cite{bv13}) on the phenomenological noise model for the punctured
surface code.  We improve upon their work by showing how to adapt simulations for the circuit-noise model and
demonstrate a simulation feature that enables simulations to run orders of magnitude faster with a caching scheme that is
naturally amenable to rare event simulations.

For the rest of this paper, we briefly introduce rare event simulation using the splitting method and describe the
common noise models for assessing QEC and fault-tolerance in Section~\ref{sec:splitting}.
In Section~\ref{sec:circuit_extension}, we work out the extension of the  Markov Chain Monte Carlo (MCMC) approach to the
circuit noise model.  The circuit noise model captures many of the effects that real quantum systems have at a
level that is more representative of hardware effects than the code capacity and phenomenological noise models. In
Section~\ref{sec:simulation}, we describe the parameters for our simulation study, which includes the case study
for the popular rotated surface codes.  We extensively show results from our simulations across a range of code distances and
provide our insights on the agreement between the Monte Carlo and rare event simulation in the accessible regime.
Finally, we conclude this work in Section~\ref{sec:future} and state the future directions for this work in
studies that have qubit leakage effects and circuits with dynamic size imposed by post-selections.

\section{Splitting Method}\label{sec:splitting}

The splitting method is a Metropolis-Hastings algorithm \cite{chib1995understanding} that can
much more efficiently explore a parameter space, which we adapt for the estimation of logical performance of QEC circuits
under circuit-noise models.  Traditionally, the assessment of logical
performance of QEC circuits relies on Monte Carlo computer simulation approaches due to the many different gates involved in a QEC circuit.  However, as the QEC circuit size increases or the physical error rates get lower, assessments using
Monte Carlo simulations start becoming
too time consuming. In fact, for low physical error rate ($p$) regimes (much below the pseudo-threshold of QEC circuits),
standard Monte Carlo sampling becomes infeasible, as the number of samples needed for an estimate becomes exorbitant.  For example,
trying to estimate logical failure rates ($\overline{p}$) in the teraquop regime would rely on order of a hundred
trillion independent runs.  For a strictly
fault-tolerant encoded circuit, $\overline{p}=\Omega\left(p^{\lceil d/2\rceil}\right)$ as $p$ approaches 0,
where $d$ is the distance of the $[[n,k,d]]$ code. The $[[n,k,d]]$ convention means a stabilizer code with $n$ physical qubits, $k$ logical qubits and minimum distance $d$.
   
To reduce the number of Monte Carlo samples needed to get an estimate, one can limit the size of gate failure sets sampled, noting that for large enough size $m$, the logical failure rate as a function of the physical error rate is:
\begin{align*}
\overline{p}&(p) \approx \\&\sum_{k=0}^{m}\binom{G}{k} p^k (1-p)^{G - k}\\&\cdot \Pr(\text{logical fail}| \text{no. failing gates}=k),
\end{align*}
where $G$ is the number of quantum gates in the encoded circuit. This technique is sometimes called subset sampling~\cite{heussen2024dynamical}.

\begin{figure}[!htbp]
\begin{minipage}{.365\columnwidth}
{\small
\begin{tabular}{|c|c|}
\hline
 & \% of $\lceil\frac{d}{2}\rceil$\\ $d$&malignant \\&faults\\\hline
3 &$5.65\times10^0$\\\hline
5&$7.90\times10^{-2}$\\\hline
7&$3.49\times10^{-4}$\\\hline
9&$6.88\times10^{-7}$\\\hline
\end{tabular}
}
\end{minipage}
\hspace{-1.25em}
\begin{minipage}{.65\columnwidth}
\begin{tikzpicture}[scale=.65]
 \begin{axis}[grid=both,xlabel = {$d$}, ylabel = {\%},ymode=log, title={\% of weight $\lceil\frac{d}{2}\rceil$ malignant faults}, legend style={at={(1,0)},anchor=south east},legend cell align = {left}]
  
    \addplot[blue, semithick, dotted,mark=*] coordinates{(3,5.65) (5,0.08) (7,0.00037) (9,6.8843e-7)};
    \end{axis}
 \end{tikzpicture}
\end{minipage}
\caption{Estimated percentage of weight $\lceil\frac{d}{2}\rceil$ faults leading to logical failure for a distance-$d$ rotated surface code.}\label{fig:percent_min_faults_failing}
\end{figure}
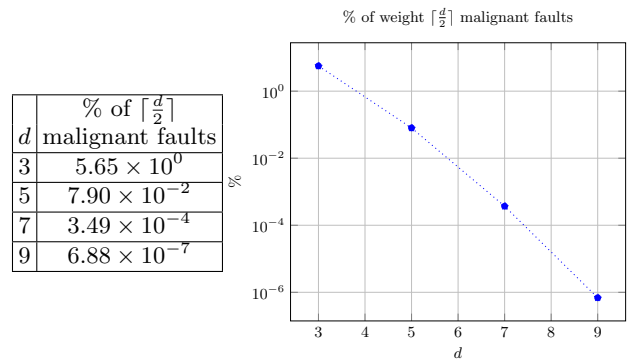

For a strictly fault-tolerant circuit, the terms with $k<\lceil \frac{d}{2}\rceil$ do not contribute to the sum as they should not lead to any
logical failures.  For the cases with large $k$, i.e., $k > m$, the $p^k$ terms become negligible and do not
contribute significantly to the sum. However, as the code distance increases, many samples are required to estimate the
$\Pr(\text{fail}| \text{no. failing gates}=k)$ term. At $k=\lceil d/2\rceil$, many of the $\binom{G}{k}$ size-$k$ sets may not lead to
failure. See Figure \ref{fig:percent_min_faults_failing} for estimates from simulation.  

Thus, estimating the probability of logical failure to sufficient accuracy given that $k=\lceil d/2\rceil$ quantum gates failed
becomes computationally challenging.  Our experience is that for large code distance, and without any guarantees of strict fault tolerance, there is a
tendency for the estimator to fail on this step because it assumes that $\Pr(\text{fail}| \text{no. failing gates}=k) = 0$, so as
not to be in an infinite loop cycle estimating what it assumes is zero.  For instance, considering a case of 10,000 faulty quantum
gates and a code distance of $d=13$,
there are potentially about $2\times10^{24}$ samples to search for failing sets of gates of size $=\lceil 13/2\rceil = 7$, which may number
in the thousands only.  That is, it would take roughly $2\times10^{20}$ samples to get a single failing event on average, which is
generally not attainable in today's classical computing systems.

The splitting method of \cite{bv13} offers an alternative, although it was limited to the \emph{code capacity} and
the \emph{phenomenological} error models (see Figure \ref{fig:error-models} for descriptions). We summarize that approach below before
extending to the more complicated \emph{circuit} model in Section \ref{sec:circuit_extension}.

\begin{figure}[!htbp]
\centering
\fbox{
\begin{minipage}{0.4\textwidth}
\raggedright
\textbf{Noise Models}
\\~\\\emph{Code capacity}: Errors can occur on data qubits. Syndrome extraction is treated as an error-free process.
\\~\\\emph{Phenomenological}: Errors can occur on data qubits.  Syndrome extraction is treated as a black box process that allows errors on the measurements.
\\~\\\emph{Circuit}: Syndrome extraction is modeled via a quantum circuit with one and two qubit gates.  Every gate is modeled as an ideal gate composed with a Pauli noise channel.  Gates fail independently, although this can be relaxed by generalizing the sampling routine.  Errors can be propagated between data and ancilla qubits during syndrome extraction, e.g., through couplings with CNOT, CPHASE, or ISWAP gates.
\end{minipage}
}
\caption{Descriptions of three common error models.}\label{fig:error-models}
\end{figure}

Consider a monotonic sequence of parameterized physical failure rates, $p \in \{p_i\}_{i=1}^t$ and denote the corresponding output failure
rates by $\{\overline{p}_i\}_{i=1}^t$. If the sequence of physical failure rates is decreasing and $\overline{p}_i$ is small, it may be
easier to estimate $\overline{p}_1$ and the sequence of values $\left\{\frac{\overline{p}_{i+1}}{\overline{p}_{i}}\right\}_{i=1}^{t-1}$
than it is to directly estimate the target output logical failure rate $\overline{p}_t$.  This initial output failure rate, $\overline{p}_1$,
can be estimated from a Monte Carlo simulation at the parameterized input rate of $p_1$ chosen to be relatively large so that Monte Carlo
can estimate in a timely manner.  In general, this splitting approach can be extended to also simulate cases where the target failure rate,
$\overline{p}_t$, we wish to estimate is larger than the initial output logical rate, $\overline{p}_1$.  However, the latter is often not
the scenario most researchers and engineers are interested in; they often desire estimates that are closer to the teraquop regime and
thus the target logical rate is much smaller than the initial failure rate estimates that Monte Carlo can achieve.

For a surface code that relies on a decoding graph, the authors prescribed a way to choose the sequence interval points for the rare event
simulation using the heuristic
\begin{equation}
    p_{i+1} = p_i\cdot 2^{\pm 1/\sqrt{w_i}},\label{eqn:heuristic}
\end{equation}  where $w_i=\max(d/2, p_i\cdot|\text{\{edges in decoding 
graph}\}|)$.  The $\pm$ sign in the exponent of Equation~\ref{eqn:heuristic} defines the direction the sequence will go as a function of where the target failure rate is
with respect to the initial output failure rate.  For most users, the intent would be to use rare event simulation to estimate the lower
target logical failure rate, so the sequence becomes: $p_{i+1} = p_i\cdot 2^{-1/\sqrt{w_i}}$.  The intuition behind choosing $w_i$ is to
start around the expected number of gates that will cause logical failures.  Thus, for any quantum circuit encoded in an error-correcting
code (not just the surface codes), we can adapt the splitting method so that $w_i = \max(d/2, p_i \cdot G)$, where $G$ is the number of
faulty gates in the circuit as defined before.  We illustrate in the Appendix that the heuristic for picking the
sequence $\left\{\frac{p_{i+1}}{p_{i}}\right\}_{i=1}^{t-1}$ is really a suggestion for getting roughly the same ratios between sequence
intervals, but other convenient choices would still lead to the same estimate of the target logical failure rate desired.

\subsection{Notation}
To describe the splitting method and prescribe the algorithm, we introduce a few notations which closely follow \cite{bv13}. Let:
\begin{itemize} 
    \item  $\Omega$: sample space of all possible \emph{events}, i.e., the powerset of all possible (gate, fault) tuples in the QEC-encoded circuit;
    \item $\mathcal{F}$: set of all failing (malignant) events, which are all the elements of $\Omega$ that individually lead to a logical failure in the QEC-encoded circuit;
    \item $\pi_{i}(E)$: probability of event $E \in \Omega$ at step $i$ in the sequence of simulated $p_i$ values; 
    \item $\pi_{i|\mathcal{F}}$ is the conditional probability distribution over failing events defined as:
    \begin{equation*} \pi_{i|\mathcal{F}} (E) := \left\{
	   \begin{array}{ll}
		\frac{\pi_{i}(E)}{\pi_{i}(\mathcal{F})}  & \mbox{if } E \in \mathcal{F} \\
		  0 & \mbox{if } E \notin \mathcal{F}
	   \end{array}
    \right.
    \end{equation*}
    \item $\mathbb{E}_{i|\mathcal{F}}(\cdot)$: expectation with respect to $p_i$ conditioned on the failure set, $\mathcal{F}$.
\end{itemize}
\subsection{Ratio Estimation}
For some function $g:\mathbb{R^+}\to\mathbb{R^+}$ such that $g(x)=x^{-1}g(x^{-1})$ and any constant $C>0$,
\begin{equation}
\frac{\overline{p}_{i+1}}{\overline{p}_i}=\frac{\pi_{i+1}(\mathcal{F})}{\pi_i(\mathcal{F})}=C\frac{\mathbb{E}_{i|\mathcal{F}}\left[g\left(C\frac{\pi_i}{\pi_{i+1}}\right)\right]}{\mathbb{E}_{i+1|\mathcal{F}}\left[g\left(C^{-1}\frac{\pi_{i+1}}{\pi_{i}}\right)\right]}.
\label{eqn:c_ratio}
\end{equation}

For a fixed number of samples, $N$, the function $g$ minimizing the statistical error is $g(x)=\frac{1}{1+x}$
\cite{bennett76}. The expectations on the numerator and denominator will be approximated with empirical averages:

\begin{align}\mathbb{E}_{i|\mathcal{F}}\left[g\left(C\frac{\pi_i}{\pi_{i+1}}\right)\right]
&\approx \mathbb{E}_{i|\mathcal{F}}^{\text{est}}\left[g\left(C\frac{\pi_i}{\pi_{i+1}}\right)\right] \nonumber \\
&:=\frac{1}{N}\sum_{j=1}^N g\left(C\frac{\pi_i(E_j)}{\pi_{i+1}(E_j)}\right),
\label{eqn:empiricalavg}
\end{align}
where the $E_1,\dots, E_N \in\mathcal{F}$ are drawn with the probability distribution $\pi_{i|\mathcal{F}}$,
found using a Metropolis routine to generate a reversible ergodic Markov chain (see \ref{sec:phenom-markov}).

Using the empirical estimates of the expectations, our goal is to find $C$ such that
\begin{equation}
\mathbb{E}_{i|\mathcal{F}}^{\text{est}}\left[g\left(C\frac{\pi_i}{\pi_{i+1}}\right)\right] = \mathbb{E}_{i+1|\mathcal{F}}^{\text{est}}\left[g\left(C^{-1}\frac{\pi_{i+1}}{\pi_{i}}\right)\right].
\label{eqn:c_satisfy}
\end{equation}
The $C$ value that results in the above equality gives an estimate of the desired ratio $\frac{\overline{p}_{i+1}}{\overline{p}_i}$.  This is apparent
from Equation~\ref{eqn:c_ratio} because the ratio of expectations becomes unity and one is left with
$\frac{\overline{p}_{i+1}}{\overline{p}_i}$ on the left-hand side and $C$ on the right-hand side.

\subsection{Reversible Markov Chain}\label{sec:phenom-markov}
The goal of the Metropolis routine is to produce for each $i$, an aperiodic and irreducible, hence ergodic, Markov Chain (RIMC) whose stationary distribution is $\pi_{i|\mathcal{F}}$.
This enables us to estimate the empirical averages on the right-hand side of Equation~\ref{eqn:empiricalavg}.
The Metropolis routine initializes a failing set $E\in\mathcal{F}$ and
uses Bennett's acceptance method \cite{bennett76} to produce a RIMC as follows. At each iteration:
\begin{enumerate}
\item Select a (gate, fault) tuple $e$ (i.e., a set of cardinality 1 in $\Omega$) uniformly at random with respect to the unit-cardinality sets, and let $E'$ be the symmetric difference of $E,e$ i.e., $E\oplus e$.
For the surface code and the phenomenological noise model, $e$ can be thought
of as an edge in the decoder graph (defined in Section~\ref{sec:circuit_extension}); this is not the case for circuit noise models, as will be seen in Section~\ref{sec:circuit_extension}.
\item Compute $q=\min\left\{1,\frac{\Pr(E')}{\Pr(E)}\right\}$ and perform a Bernoulli trial with success probability $q$. This is the
acceptance step.
\item If the Bernoulli trial resulted in success and $E'\in\mathcal{F}$, output $E'$. Otherwise, output $E$.
\end{enumerate}
The routine results in a Markov Chain that satisfies the detailed balance equation: \begin{align}
    \frac{\Pr(E')}{\Pr(E)} & = \frac{\Pr(E'|E)}{\Pr(E|E')} \label{eqn:DBE}
\end{align}
and the forward chain, $E\rightarrow E'$, has the same probability as with the backward chain, $E'\rightarrow E$.  Furthermore the chain is irreducible since one can start with a failing set $E$ and arrive at another $E'$ using the elements of $E \oplus E'$. So the Markov chain is a RIMC and its probability distribution will converge to $\pi_{i|\mathcal{F}}$.

\section{Extension to Circuits}\label{sec:circuit_extension}

When using the circuit noise model, (gate, fault) pairs can introduce errors to the data qubits and the syndrome measurements.  There exists a class of graph-based decoders such as the \emph{minimum weight perfect matching} (MWPM) and \emph{union-find} (UF) for surface codes that can be described on a weighted graph that we call the decoder graph.  The vertices of the decoder graph are the $X$ ($Z$) checks and the edges represent gate-faults or errors that result in the incident checks changing values.  In the phenomenological noise model, selecting errors simply corresponds to selecting edges in the decoder graph.  In the circuit noise model multiple gate-faults map to the same syndrome (edges).  Due to this nature of circuit noise, we modify
the state space of the MC in the original protocol in \cite{bv13} from sets of edges to gate-faults.    When doing so, we must modify the Metropolis routine from selecting edges in the decoding graph to selecting physical (gate, fault) pairs.  Our Metropolis routine selection protocol results in a reversible MC.  This is because as we show below the MC transition probabilities satisfy the detailed balance equations. 
 
Starting with a set $E$ of (gate, fault) pairs leading to logical failure, we select a (gate, fault) pair $(g,f)$ uniformly at random in
the circuit. If $g$ is not in any (gate, fault) pair in $E$ regardless of the fault, then set $E'=E\cup\{(g,f)\}$. Otherwise, $(g,h) \in E$ for some
fault $h$. In the latter case, set
$E'=(E \cup \{(g,f)\})\setminus\{(g,h)\}$. Note that if the faults $f$ and $h$ are the same in this latter case,
then  $|E'| = |E|-1$, where the $(g,f)$ has been dropped.  The selection process ensures $E'$ is physically viable as opposed to a symmetric difference which would allow a single gate with two different faults.
For $E'\in\mathcal{F}$,  the acceptance probability, $\Pr(E'|E)$, is chosen as follows:
    \begin{itemize}
        \item \textbf{Case 1}:  $g$ is not in any gate fault pair in $E$. Set the acceptance probability to \[\frac{\Pr(g)}{1-\Pr(g)}\cdot \textstyle\Pr_g(f),\]
        where $\Pr(g)$ is the probability of gate $g$ failing and $\Pr_g(f)$ is the probability of the fault $f$ occurring given that $g$ failed.
        \item \textbf{Case 2}: There exists a fault $h$ such that $(g,h)\in E$.
        \begin{enumerate}
            \item If $f=h$, set the acceptance probability to $1$.
            \item If $f\neq h$, set the acceptance probability to $\Pr_g(f)$.
        \end{enumerate}
    \end{itemize}
In each case, the detailed balance equation is satisfied:
    \begin{itemize}
      \item \textbf{Case 1}: $g$ is not in any gate fault pair in $E$, $E' = E\cup\{(g,f)\}$.
      \begin{align*}&\frac{\Pr(E')}{\Pr(E)} \\&= \frac{\prod_{(i,j)\in E'}\left(\Pr(i)\Pr_i(j)\right)\prod_{(i,j)\notin E'}\left(1-\Pr(i)\right)}{\prod_{(i,j)\in E}\left(\Pr(i)\Pr_i(j)\right)\prod_{(i,j)\notin E}\left(1-\Pr(i)\right)} \\
      &= \frac{\Pr(g)\Pr_g(f)}{1-\Pr(g)}\\
      &= \frac{\frac{\Pr(g)}{1-\Pr(g)}\cdot \Pr_g(f)}{1}\\
      &=\frac{\Pr(E'|E)}{\Pr(E|E')}.  \end{align*} 
        \item \textbf{Case 2}: There exists a fault $h$ such that $(g,h)\in E$.
        \begin{enumerate}
            \item If $f=h$, $E' = E\setminus\{(g,f)\}$  or equivalently $E = E' \cup \{(g,f)\}$.  This case is the reciprocal of Case 1.
            \item If $f\neq h$:
            \begin{align*}\frac{\Pr(E')}{\Pr(E)} = \frac{\Pr_g(f)}{\Pr_g(h)} =
      \frac{\Pr(E'|E)}{\Pr(E|E')}. \end{align*} 
        \end{enumerate}
    \end{itemize}
Note that as stated, these acceptance probabilities apply to modifying $E$ by a single fault to get $E'$.

\subsection{Extension to Leakage}\label{sec:leakage}

The approach we have for the circuit model with Pauli noise extends to leakage naturally.  The only difference is
that in addition to faults we include the leakage paths.  Leakage paths are the set of errors induced on the
qubits that interact with a leaked qubit.  The states in the Markov chain become the sets of (gate, fault, path) leading to logical
failure.  We select a tuple $e:=(g,f,\rho)$ uniformly at random from the set of triples in the circuit.  Note a
path $\rho$ could be empty if, for instance, the gate fault does not induce leakage, i.e., the fault is purely a
Pauli error. 
We now describe the construction of the new malignant set $E'$ from a malignant set $E$ and setting of $q$, the
acceptance probability, as $\min\left\{1,\frac{\Pr(E')}{\Pr(E)}\right\}$. If $g$ is not in any gate in $E$ regardless of the fault, then set $E'=E\cup e$. Otherwise, $(g,h,\nu) \in E$ for some 
fault $h$ and an optional leakage path $\nu$. In the latter case, set
$E'=(E \cup \{(g,f,\rho)\})\setminus\{(g,h,\nu)\}$. Note that if $(f,\rho)=(h,\nu)$ then  $|E'| = |E|-1$, where 
the $(g,f, \rho)$ has been dropped.  The selection process ensures $E'$ is physically viable as opposed to a 
symmetric difference which would allow a single gate with two different faults, an unphysical occurrence.

We consider three cases: 
\begin{itemize}
    \item \textbf{Case 1}:  
For $E'\in\mathcal{F}$,  the acceptance probability, $\Pr(E'|E)$, is chosen as follows:
    $g$ is not in any gate fault pair in $E$. Set the acceptance probability to \[\frac{\Pr(g)}{1-\Pr(g)}\cdot \textstyle\Pr_g(f) \Pr_f(\rho),\]
        where $\Pr(g)$ is the probability of gate $g$ failing and $\Pr_g(f) \Pr_f(\rho)$ is the probability of the fault $f$ occurring given that $g$ failed.
        \item \textbf{Case 2}: There exists a fault $h$ such that $(g,h, \nu)\in E$.
        \begin{enumerate}
            \item If $(f, \rho) =(h, \nu)$, set the acceptance probability to $1$.
            \item If $(f, \rho) \neq (h, \nu)$, set the acceptance probability to $\Pr_g(f) \Pr_f(\nu)$.
        \end{enumerate}
    \end{itemize}
One can follow the analysis in the non-leakage case and show that the detailed balance equation is satisfied.
Note that the addition of leakage can cause a gate selection probability to significantly increase in the
proposal step.  However, the acceptance step ensures that the total probability is proportional to $\Pr(g)$.  For
example suppose $g_1$ has a single fault $f_1$ with no fault paths due to leakage and $g_2$ has a single fault
$f_2$ and $M$ paths due to leakage.  The probability of adding $g_1$ to a malignant set $E$ not containing $g_1$ is:
\[ \frac{1}{N} \frac{\Pr(g_1)}{1-\Pr(g_1)}, \]
where $N$ is the number of triples (gate, fault, path) that can occur in the circuit.
On the other hand, the probability of adding $g_2$ to a malignant set $E$ not containing $g_2$ is:
\begin{align*}
\frac{1}{N} & \frac{\Pr(g_2) \sum_\rho \Pr_{f_2}(\rho)}{1-\Pr(g_2)}
\end{align*}
which, since $\sum_\rho \Pr_{f_2}(\rho) = 1$, is equivalent to
\begin{align*}
      & \frac{1}{N} \frac{\Pr(g_2)}{1-\Pr(g_2)}
\end{align*}
So the ratio of selecting $g_1$ versus $g_2$ remains unperturbed due to leakage.  What is different due to leakage is the increase in the state space which results in the increase in $N$ and hence the number of trials required for selecting a specific $e$.

\subsection{Extensions to Post-selection}\label{sec:postselection}
So far our Metropolis routine alters the state space by a single element.  The drawback with this approach is that in the context of post-selection or logical circuits with multiple qubits, it will not necessarily lead to ergodic Markov chains.  Consider an example where a quantum state is prepared using post-selection a typical approach in magic state cultivation and distillation~\cite{li2015magic, gidney2024magic, sales2025experimental}.  If we start with a malignant set $E$, then the gate faults of $E$ must lead to a zero syndrome in the post-selection part of the circuit.  However, altering $E$ by one within the ancilla state preparation and checks part of the circuit will typically lead to a non-zero syndrome and will no longer be malignant. Hence our Markov chain will lead to a collection of islands (absorption states) when we allow only single gate alterations in the Metropolis routine.  In the following we present a Metropolis method that allows for multi-gate alterations and results in a reversible Markov chain.  We describe the process in the context of no-leakage to simplify the notation and exposition.  However, this can be easily extended to the case of leakage by adding fault-paths to the gate-faults.  This comes at a higher computational complexity that grows linearly with the multi-gate selection parameter $k$.

We assume the Metropolis routine is initialized with a malignant set of gate-faults $E$.
\begin{itemize}
\item{\bf Proposal Step:} Our sampling set contains sets of gate-faults up to cardinality $k$, i.e., $$ S:= \{ \{(g_1, f_1)\}, \ldots, \{(g_1,f_1), \ldots, (g_k, f_k) \} \}$$ with distinct gates in the member sets that is sampled uniformly.  The sampled outcome, $\Delta$, will be used to modify the current malignant set $E$ to form $E'$.  Let $k_1:= |\Delta-E|$ be the number of additions, $k_2 := |\Delta \cap E |$ be the number of deletions and $k_3$ be the number of altered gates in $\Delta$, i.e., they exist in $E$ and have different faults.  Note that $k=k_1+k_2+k_3$.
\item {\bf Acceptance Step:}\\  Without loss of generality, assume the gate-fault additions are $(g_1, f_1)\ldots (g_{k_1}, f_{k_1})$, the deletions are $(g_{k_1+1}, f_{k_1+1}) \ldots (g_{k_1+k_2}, f_{k_1+k_2})$, and the alterations in $\Delta$ are $(g_{k-k_3+1}, f_{k-k_3+1}) \ldots (g_k, f_k)$.  We compute the Bernoulli acceptance parameter 
\begin{align*} 
& p_1 := \textstyle\frac{\Pr(g_1)}{1-\Pr(g_1)} \textstyle\Pr_{g_1}(f_1) \dots \frac{\Pr(g_{k_1})}{1-\Pr(g_{k_1})} \textstyle\Pr_{g_{k_1}}(f_{k_1}) \\
& p_2:= 1\\
& p_3 :=  \textstyle\Pr_{g_{k-k_3+1}}(f_{k-k_3+1}) \ldots \textstyle\Pr_{g_k}(f_k) \\
& p := p_1 p_2 p_3
\end{align*}
where $\Pr(g_i)$ is the probability of gate $g_i$ failing and $\Pr_{g_i}(f_i)$ is the probability of the fault $f_i$ occurring given that $g_i$ failed.
\end{itemize}

 Recall we need to show that the above Metropolis routine satisfies the Detailed Balance Equation (\ref{eqn:DBE}). 
 Assume $E,E'$ and $k,k_1,k_2,k_3$ are as defined above.  Note that the additions for the $E \rightarrow E'$ transitions become subtractions for the $E' \rightarrow E$, and vice versa.  The alterations for $E \rightarrow E'$ and $E' \rightarrow E$ are the same gates, but the faults swap from $f_i$ to $h_i$.  Since the gate-faults are an independent process we have
 \begin{align*}
 &\frac{\Pr(E')}{\Pr(E)} =  \\& \frac{\prod_{i=1}^{k_1}\frac{\Pr(g_i)}{1-\Pr(g_i)}\Pr_{g_i}(f_i)}{\prod_{i=k_1+1}^{k_1+k_2}\frac{\Pr(g_i)}{1-\Pr(g_i)}\Pr_{g_i}(f_i)}\times \prod_{i=k-k_3+1}^{k} \frac{\Pr_{g_i}(f_i)}{\Pr_{g_i}(h_i)}\\
 &= \frac{\Pr(E'|E)}{\Pr(E|E')}\label{eqn:kDBE}
 \end{align*}
 where the last equality follows from the definition of the Metropolis routine in the acceptance step.

\section{Simulation Setup}\label{sec:simulation}

Here, we describe the setup for the simulation studies used in our analysis of the
rare event error model.  Our simulation studies were conducted with the $[[d^2, 1, d]]$
rotated surface code.  This code is widely studied~\cite{PhysRevA.86.032324, dennis2002topological,google2023suppressing,horsman2012surface}
for its many advantages
towards a scalable quantum computer, namely locality, simplicity of the syndrome extraction,
and the existence of efficient decoder families.  In our studies, we use the efficient MWPM decoder but note that other efficient decoders like
the \emph{Union-Find} decoder \cite{delfosse2021almost, Delfosse2021unionfinddecoders} can be used as well.  While our studies were for the rotated surface code under
the MWPM decoder, the MCMC piece of our simulation has no real dependence on decoder selection
and, as such, extends to other code and decoder configurations of interest.

The smallest error-correcting member of the medial surface code can be described as:
\begin{figure}[!htbp]
\centering
\begin{tikzpicture}[scale=1]

\fill[blue!30](0,0)--(0,1)--(1,1)--(1,0)--(0,0){};
\fill[red!30](1,0)--(1,1)--(2,1)--(2,0)--(1,0){};
\fill[blue!30](1,-1)--(1,0)--(2,0)--(2,-1)--(1,-1){};
\fill[red!30](0,-1)--(0,0)--(1,0)--(1,-1)--(0,-1){};

\fill[red!30](0,1) to [bend left, in=90,out=90,looseness=1.5] (1,1){};
\fill[blue!30](2,1) to [bend left, in=90,out=90,looseness=1.5] (2,0){};
\fill[red!30](2,-1) to [bend left, in=90,out=90,looseness=1.5] (1,-1){};
\fill[blue!30](0,-1) to [bend left, in=90,out=90,looseness=1.5] (0,0){};

\draw[very thin](0,1)--(2,1)--(2,-1)--(0,-1)--(0,1){};
\draw[very thin](0,1) to [bend left, in=90,out=90,looseness=1.5] (1,1) --(1,-1){};
\draw[very thin](2,1) to [bend left, in=90,out=90,looseness=1.5] (2,0){};
\draw[very thin](2,-1) to [bend left, in=90,out=90,looseness=1.5] (1,-1){};
\draw[very thin](0,-1) to [bend left, in=90,out=90,looseness=1.5] (0,0)--(2,0){};
\end{tikzpicture}

\caption{A distance-3 medial surface code, with \textcolor{red}{$X$} checks and \textcolor{blue}{$Z$} checks. Physical data qubits are at the corners.  Each color face is a check with qubits incident.}
\label{fig:scd3}
\end{figure}
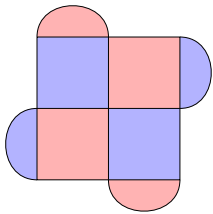
This code consists of $X$ checks and $Z$ checks as represented by the checkered pattern of red and blue.
Figure~\ref{fig:scd3} is the $[[9,1,3]]$ member of the family of medial surface code because it has 9 data
qubits distributed at the corners of the squares and it can fix any $\frac{d-1}{2}=1$ error.  The other members of
this medial surface code follow the same pattern with the grid that is required to yield a
$d^2$ layout of data qubits, where $d$ is the code's distance.  In the circuit realization of this code, there is a
need to follow a certain routine for correctness and fault tolerance of syndrome extraction.  In particular, we choose a syndrome extraction
routine that does not permit single faults to spread maliciously as shown in Figures~\ref{fig:xstab} and
\ref{fig:zstab}.
\begin{figure}[!htbp]
  \begin{minipage}{0.3\textwidth}
  \centering
    \begin{tikzpicture}[thick,scale=.7]

    \tikzstyle{operator} = [draw,fill=white,minimum size=1.5em]
    \tikzstyle{control} = [fill,shape=circle,minimum size=5pt,inner sep=0pt]
    \tikzstyle{phase} = [fill,shape=circle,minimum size=5pt,inner sep=0pt]
    \tikzstyle{not} = [draw,fill=white,shape=circle,minimum size=5pt,inner sep=-1pt]
    \tikzstyle{surround} = [fill=red!10,thick,draw=black,rounded corners=2mm]

    \node at (0,0) (q1) {\ket{+}};
    \node at (0,-1) (q2) {\ket{d}$_{\mbox{\tiny NW}}$};
    \node at (0,-2) (q3) {\ket{d}$_{\mbox{\tiny NE}}$};
    \node at (0,-3) (q4) {\ket{d}$_{\mbox{\tiny SW}}$};
    \node at (0,-4) (q5) {\ket{d}$_{\mbox{\tiny SE}}$};

    \node[control] (op2) at (1,0) {} edge [-] (q1);
    \node[not] (op3) at (1,-1) {\textbf{+}} edge [-] (q2);
    \draw[-] (op2) -- (op3);

    \node[control] (op4) at (2,0) {} edge [-] (op2);
    \node[not] (op5) at (2,-2) {\textbf{+}} edge [-] (q3);
    \draw[-] (op4) -- (op5);

    \node[control] (op6) at (3,0) {} edge [-] (op4);
    \node[not] (op7) at (3,-3) {\textbf{+}} edge [-] (q4);
    \draw[-] (op6) -- (op7);

    \node[control] (op8) at (4,0) {} edge [-] (op6);
    \node[not] (op9) at (4,-4) {\textbf{+}} edge [-] (q5);
    \draw[-] (op8) -- (op9);

    \node[operator] (op11) at (5,0) {$M_X$} edge [-] (op8);

    \node (end1) at (6,0) {} edge [-] (op11);
    \node (end2) at (6,-1) {} edge [-] (op3);
    \node (end3) at (6,-2) {} edge [-] (op5);
    \node (end4) at (6,-3) {} edge [-] (op7);
    \node (end5) at (6,-4) {} edge [-] (op9);

    \begin{pgfonlayer}{background} 
    \node[surround] (background) [fit = (q1) (op9) (end5)] {};
    \end{pgfonlayer}

    \end{tikzpicture}
    \end{minipage}
    \begin{minipage}{0.15\textwidth}
    \begin{tikzpicture}
    \fill[red!10] (-0.25,-0.25) rectangle (2.25,2.25);
    
    \draw[thick, red] (0,2) -- (2,2) -- (0,0) -- (2,0);
    
    \draw[->, thick, red] (0,0) -- (2,0);
    \end{tikzpicture}
    \end{minipage}
    
    \caption{CNOT couplings for measuring a $X$-type stabilizer generator in the standard gate basis along with the direction of coupling operations on an \textcolor{red}{$X$}-type plaquette of a surface code patch. The subscript of NW, SW, NE, and SE on the data qubit represent
the northwest, southwest, northeast, and southeast locations of the data with respect to the $X$ ancilla in the grid layout.}
    \label{fig:xstab}
\end{figure}
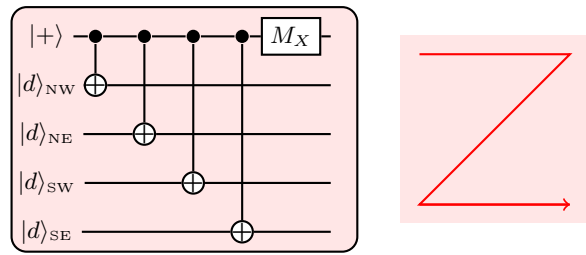

\begin{figure}[!htbp]
  \begin{minipage}{0.3\textwidth}
  \centering
    \begin{tikzpicture}[thick,scale=.7]

    \tikzstyle{operator} = [draw,fill=white,minimum size=1.5em]
    \tikzstyle{control} = [fill,shape=circle,minimum size=5pt,inner sep=0pt]
    \tikzstyle{phase} = [fill,shape=circle,minimum size=5pt,inner sep=0pt]
    \tikzstyle{not} = [draw,fill=white,shape=circle,minimum size=5pt,inner sep=-1pt]
    \tikzstyle{surround} = [fill=blue!10,thick,draw=black,rounded corners=2mm]

    \node at (0,0) (q1) {\ket{0}};
    \node at (0,-1) (q2) {\ket{d}$_{\mbox{\tiny NW}}$};
    \node at (0,-2) (q3) {\ket{d}$_{\mbox{\tiny SW}}$};
    \node at (0,-3) (q4) {\ket{d}$_{\mbox{\tiny NE}}$};
    \node at (0,-4) (q5) {\ket{d}$_{\mbox{\tiny SE}}$};

    \node[not] (op2) at (1,0) {\textbf{+}} edge [-] (q1);
    \node[control] (op3) at (1,-1) {} edge [-] (q2);
    \draw[-] (op2) -- (op3);

    \node[not] (op4) at (2,0) {\textbf{+}} edge [-] (op2);
    \node[control] (op5) at (2,-2) {} edge [-] (q3);
    \draw[-] (op4) -- (op5);

    \node[not] (op6) at (3,0) {\textbf{+}} edge [-] (op4);
    \node[control] (op7) at (3,-3) {} edge [-] (q4);
    \draw[-] (op6) -- (op7);

    \node[not] (op8) at (4,0) {\textbf{+}} edge [-] (op6);
    \node[control] (op9) at (4,-4) {} edge [-] (q5);
    \draw[-] (op8) -- (op9);

    \node[operator] (op11) at (5,0) {$M_Z$} edge [-] (op8);

    \node (end1) at (6,0) {} edge [-] (op11);
    \node (end2) at (6,-1) {} edge [-] (op3);
    \node (end3) at (6,-2) {} edge [-] (op5);
    \node (end4) at (6,-3) {} edge [-] (op7);
    \node (end5) at (6,-4) {} edge [-] (op9);

    \begin{pgfonlayer}{background} 
    \node[surround] (background) [fit = (q1) (op9) (end5)] {};
    \end{pgfonlayer}

    \end{tikzpicture}
    \end{minipage}
    \begin{minipage}{0.15\textwidth}
    \begin{tikzpicture}
    \fill[blue!10] (-0.25,-0.25) rectangle (2.25,2.25);
    
    \draw[thick, blue] (0,2) -- (0,0) -- (2,2) -- (2,0);
    
    \draw[->, thick, blue] (2,2) -- (2,0);
    \end{tikzpicture}
    \end{minipage}
    
    \caption{CNOT couplings for measuring a $Z$-type stabilizer generator in the standard gate basis along with the direction of coupling operations on a \textcolor{blue}{$Z$}-type plaquette of a surface code patch. The subscript of NW, SW, NE, and SE on the data qubit represent
the northwest, southwest, northeast, and southeast locations of the data with respect to the $Z$ ancilla in the grid layout.}
    \label{fig:zstab}
\end{figure}

The circuits above are repeated at each corresponding site for the $X$ and $Z$ syndrome bits, then repeated
over $r=2d$ rounds of syndrome extraction for fault-tolerance to the distance of the given code.  There are
a number of simulation software for studying the rotated surface code, including Stim~\cite{gidney2021stim}.
In our studies, we perform simulation and evaluation of syndrome extraction in the following way:
\begin{enumerate}
    \item For a $[[d^2,1,d]]$ surface code, we perform noisy syndrome extractions for $2d$ rounds.
    \item We perform error correction only within the first $d$ rounds.
\end{enumerate}
This approach simulates logical idle and differs from ``logical T1'' simulation setup of Stim.
The choice of applying error correction only on the first $d$ rounds is
motivated by the need to be strictly fault-tolerant.  Intuitively, having a buffer of extra $d$ rounds
enables the user to distinguish between data errors and measurement errors.  The syndrome information
embedded in the latter $d$ rounds can then be used for correction after an extra $d$ rounds of syndrome
extraction.

In our circuit noise model, every quantum gate, including preparations and measurement,
is assumed to have a failure rate parameterized by $p$, the physical gate failure rate.  For our
simulation results, unless noted, all gates fail with probability $p$.  It is possible to extend to the so-called \texttt{Pauli+} noise model, where
gates are able to cause leakage on the input qubit(s); we do not include the \texttt{Pauli+} noise model in our study.

We perform simulations in the accessible regime of standard Monte Carlo, and rare event simulations in the accessible regime and well-beyond the accessible regime of the Monte Carlo runs.  The
results from these scenarios enable us to assess the correctness of the rare event simulation
technique in the regions of overlap, but then shows us results for the overall logical idle
simulations, across all $2d$ rounds of syndrome extraction in the teraquop range.  

A Monte Carlo simulation at the initial (highest) physical failure rate $p_1$ provides an estimate for $\overline{p}_1$ as well as a starting event for the Markov chain. We save the first malignant set (causing logical failure) during the Monte Carlo simulation as the initial event $E_0$. For $i>1$, the first event we use for $p_i$ is the last event from the chain for $p_{i-1}$.

The largest 
runtime cost for these simulations tends to be the calls to the decoder. The Metropolis routine results in exploring the same set of malignant sets multiple times, hence caching the decoding outcome reducing repetitions of the same decoding computations, which speeds up the rare event simulation.
We use a cache of recent events, containing a map from set of gate faults to the decoding outcome (i.e., failure or success). For the runs in Section~\ref{sec:results}, we use a cache size of 10M.

While a detailed study of the convergence of the Markov chains used within the rare event simulation is left as a future work, we use certain heuristics to test for convergence of the righthand side of  Equation (\ref{eqn:empiricalavg}). In particular, for a run with $N$ processes running their own Markov chains, we require that a process completes the following in order to provide an estimate:
(1) After a user-specified minimum number $J$ of jumps (number of times a new event is output in the Markov chain), discard the oldest event in the chain and continue for at least $10\%$ more jumps. The rationale for discarding the older events is to account for the burn-in process.
(2) Continue at least until the number of distinct faults across all the events in the chain (including discarded events) exceeds $|E_0|+\frac{d+1}{2}$. 

Each process continues sampling until at least $90\%$ of the processes ($=0.9N$) have met both conditions, and our simulation reports the average from those completed processes. Unless otherwise specified, the results in Section~\ref{sec:results} use $N=20$ processes and $J=10$ jumps.

\section{Results}\label{sec:results}

We present results highlighting the
strengths of the rare event simulation techniques along with
empirical evidence towards correctness by comparing to simple
Monte Carlo simulations in accessible study regimes.  We also
present results to demonstrate the properties of our rare
event simulation under symmetric and asymmetric circuit noise conditions,
as well as under different conditions for choosing the interval
points for the parameterized physical failing rates, $p$.  The setup for the simulation results presented here follow our descriptions in Section~\ref{sec:simulation}. The simulations in this section were generated on an Intel Xeon server with 96 CPUs @2.7GHz and 384GB memory. The simulations were done without leakage or multi-gate fault selection.

\begin{figure}[!htbp]

\centering
\includegraphics[width=\columnwidth]{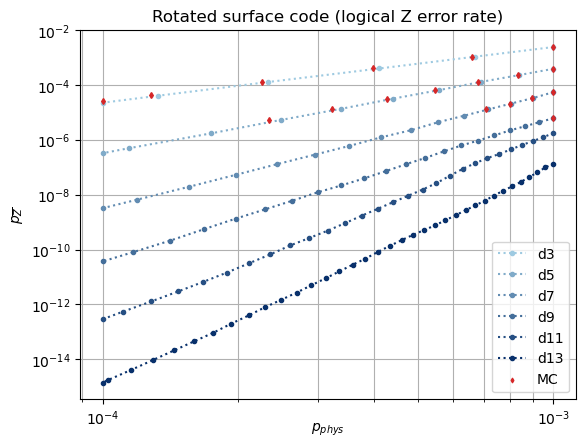}

\caption{Logical $Z$ error rate for a rotated surface code for $p_{phys}\in [10^{-4},10^{-3}]$. Estimation using Rare Event Simulation. The red diamonds show estimates using raw Monte Carlo (MC) simulation with the number of runs needed to reach at least 1,000 failures, where the estimator is the
unbiased negative binomial estimator. }\label{fig:logZerror}
\end{figure}
\begin{figure}[!ht]

\centering
\includegraphics[width=\columnwidth]{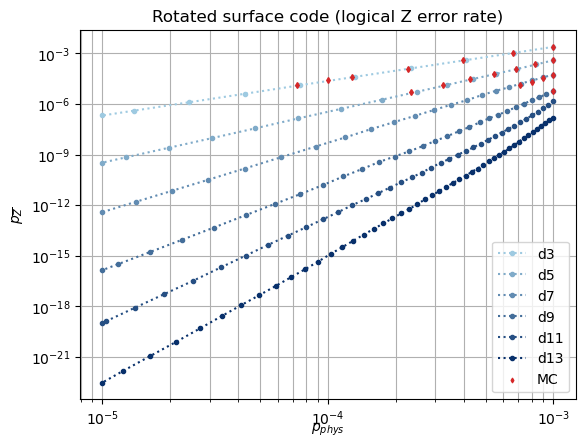}

\caption{Logical $Z$ error rate for a rotated surface code for $p_{phys}\in [10^{-5},10^{-3}]$. Estimation using rare event simulation. The red diamonds show estimates using Monte Carlo  (MC) simulation with at least 1,000 failures.}\label{fig:logZlowerp}
\end{figure}

\begin{figure}[!htbp]
\centering
\includegraphics[width=\columnwidth]{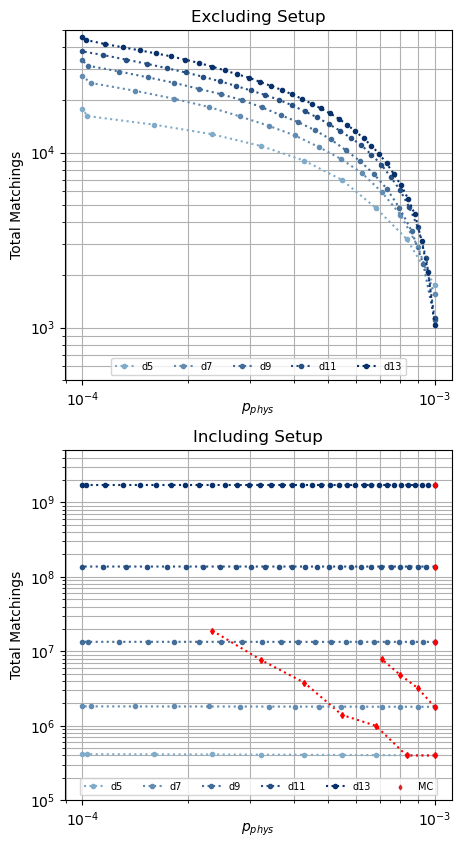}
\caption{Total Cumulative Matchings for $p_{\overline{Z}} $ estimate (using cache). (above) Excluding the setup in which initial
events and $p_{\overline{X}}$ and $p_{\overline{Z}}$ are obtained using a Monte Carlo simulation at $p_{phys}=10^{-3}$ or (below) including the setup.
The \textcolor{red}{$\blacklozenge$} points show numbers of matchings to get an estimate using Monte Carlo simulation with at least 100 failures at
a specified $p_{phys}$ (not accumulated across different $p_{phys}$).} \label{fig:matchings}
\end{figure}

We start by presenting the simulation results for the rotated
surface code in Figure~\ref{fig:logZerror} under symmetric
circuit noise conditions, that is, all quantum gates in the
syndrome extraction circuits have the same failure probability $p$. Figure \ref{fig:logZerror} shows the logical
$Z$ error rate ($p_{\overline{Z}}$) computed using rare event simulation compared 
to using Monte Carlo simulations with $p \in [10^{-4},10^{-3}]$.
These simulations are also the basis for the results shown in
Figure \ref{fig:mpi-times} and Figure \ref{fig:matchings}.
The rare event simulation can still be relied on to generate
valid results at lower $p$ as shown in Figure \ref{fig:logZlowerp}.
However, this would be practically impossible to generate
using a standard Monte Carlo approach.  For the data points
that were generated for the Monte Carlo in Figures~\ref{fig:logZerror} and \ref{fig:logZlowerp}, the results show agreement with the rare event simulation.

\begin{figure}[!htbp]
    \centering
    \includegraphics[width=\columnwidth]{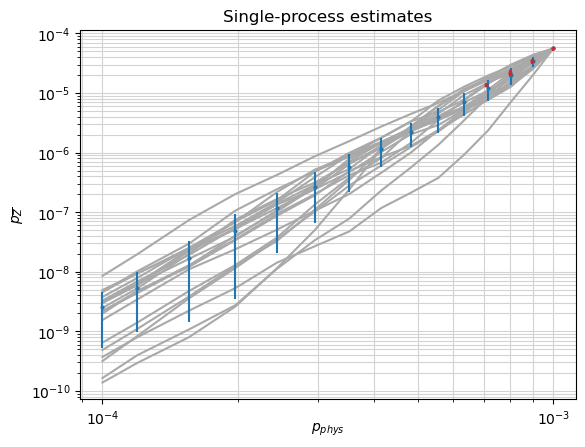}
    \caption{Logical $Z$ error rate estimates in 20 rare event simulation runs with a single process for $d=7$. The blue bars show the mean $\pm$ one standard deviation. The mean $\pm$ one standard deviation for 10 Monte Carlo runs with at least 100 failures each is shown in red.}
    \label{fig:single-run-spread}
\end{figure}

While quantification of confidence for the rare event simulation estimates is left to future work, we give an empirical treatment to convergence in Figure~\ref{fig:single-run-spread} similar to a concurrent, independent work by Beverland et al~\cite{beverland2025fail}.  Figure~\ref{fig:single-run-spread} shows the mean and standard deviation of estimates from 20 independent runs for a distance 7 rotated surface code. The logical failure estimates at each physical failure rate depend on the estimates from every previous $p_i$, so the estimation error can compound for lower physical error rates.

An interesting phenomenon with rare event simulation is that as the size of the code increases, the number of $p$ values in the heuristically-generated sequence (Equation~\ref{eqn:heuristic})
for a fixed interval of interest for $p_{phys}$ also increases.  While this potentially increases the runtime,
it is generally negligible in comparison to the
Monte Carlo setup time for the initial result at $p_0$.
Moreover, it turns out that strict adherence to the heuristically-generated sequence is not necessary. See the Appendix for further discussion.

Performing decoding, even with the MWPM decoder, is a
relatively expensive step for the failure rate estimation. 
Compared to Monte Carlo simulations in which each sample is checked for failure, the rare event simulation relies on many fewer calls to the decoder. Figure \ref{fig:matchings} shows the cumulative number of calls to a matching (MWPM) decoder to estimate the logical failure rate for a rotated surface code at $p$ in $[10^{-4},10^{-3}]$. See the Appendix for example runtimes.

The rare event simulation starts with a Monte Carlo simulation for $p=10^{-3}$, and successive smaller physical error rates use the estimate from the previous (larger) physical error rate.
We note that the frequency with which events in $\Omega$ of different cardinality are visited in the Markov chain changes with  $p$. This is illustrated in Figure \ref{fig:histogram-of-faults}. At lower $p$, a larger portion of the events have cardinality $\left\lceil\frac{d}{2}\right\rceil$.  This behavior is expected due to the fact
that the expected number of failing gates goes as $G\times p$, where $G$ is the number of quantum gates in the
encoded circuit.  However,
when $p$ becomes smaller, this expected number of failing gates is below the correction power of the code,
$\lceil \frac{d}{2}\rceil$, so the rare event simulation steadies
around events with cardinality of $\lceil \frac{d}{2}\rceil$.

\begin{figure}[!ht]
\begin{tikzpicture}[scale=.79]
    \begin{axis}[ybar interval, xlabel = {Fault set size}, ylabel = {Count}, title={d7, $\overline{Z}$, $p_{phys}=10^{-3}$, }]
    \addplot coordinates{(4,0)(5,1) (6,4) (7,17) (8,39) (9,36) (10,32) (11,27) (12,21) (13,10) (14,12) (15,2)(16,0)};
    \end{axis}
\end{tikzpicture}\\
\begin{tikzpicture}[scale=.79]
    \begin{axis}[ybar interval, xlabel = {Fault set size}, ylabel = {Count}, title={d7, $\overline{Z}$, $p_{phys}=10^{-4}$, }]
    \addplot coordinates{(4,108) (5,64) (6,13) (7,11) (8,4) (9,0)(10,0)(11,0)(12,0)(13,0)(14,0)(15,0)(16,0)};
    \end{axis}
\end{tikzpicture}
\caption{Count histograms of sizes of fault sets visited in estimation of $p_{\overline{Z}}$ at (top) $p_{phys} = 10^{-3}$ and (bottom) $p_{phys} = 10^{-4}$ for a distance-7 surface code. Note that at the lower $p_{phys}$, faults close to size $\left\lceil\frac{d}{2}\right\rceil$ account for a larger portion of events. }\label{fig:histogram-of-faults}
\end{figure}

\begin{figure}[!ht]
\centering
\begin{subfigure}[t]{\columnwidth}
\centering
\begin{tikzpicture}[scale=.25]
\fill[blue!30](0,0)--(10,0)--(10,-10)--(0,-10)--cycle{};
\foreach \r in {0,2,...,8}{
    \foreach \c in {1,3,...,9}{
        \fill[red!30] (\c,-\r)-- (\c+1,-\r)-- (\c+1,-\r-1)--(\c,-\r-1)--cycle{};
        \fill[red!30] (\c-1,-\r-1)-- (\c,-\r-1)-- (\c,-\r-2)--(\c-1,-\r-2)--cycle{};
    }    
}
\foreach \c in {0,2,...,8}{
    \fill[red!30](\c,0) to [bend left, in=90,out=90,looseness=1.5] (\c+1,0){};
    \fill[red!30](\c+2,-10) to [bend left, in=90,out=90,looseness=1.5] (\c+1,-10){};
  }
\foreach \r in {0,2,...,8}{
    \fill[blue!30](10,-\r) to [bend left, in=90,out=90,looseness=1.5] (10,-\r-1){};
    \fill[blue!30](0,-\r-2) to [bend left, in=90,out=90,looseness=1.5] (0,-\r-1){};
  }
\draw[thick](4,0.5)--(5,0.5)--(5,-10.5)--(4,-10.5)--cycle{};
\fill[pattern = {north east lines}](4,0.5)--(5,0.5)--(5,-10.5)--(4,-10.5)--cycle{};
\end{tikzpicture}
    \caption{}
\end{subfigure}
\\
\begin{subfigure}[t]{.9\columnwidth}
\includegraphics[width=\textwidth]{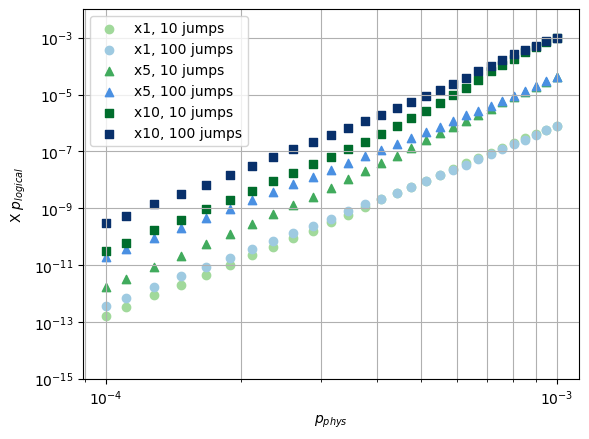}
    \caption{}
\end{subfigure}
\\
\begin{subfigure}[t]{.9\columnwidth}
\includegraphics[width=\textwidth]{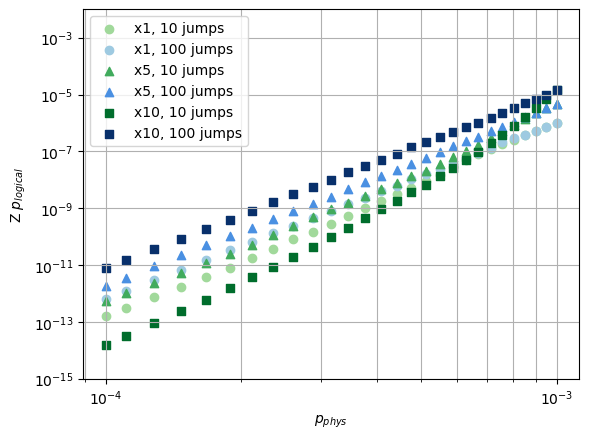}
        \caption{}
    \end{subfigure}
\caption{Logical $X$ and logical $Z$ error rates for a distance-11 rotated surface code with asymmetric noise using a minimum of 100 jumps (number of times a new event is output in the Markov chain) or 10 jumps in at least 18 of 20 processes before moving to the next $p_{phys}$. The physical error in the column marked in (a) is multiplied by the value indicated in the legend. While using 10 jumps returns results faster, doing so may not provide sufficiently accurate results.}\label{fig:asymmetric_plot}
\end{figure}

\begin{figure}
\includegraphics[width=\columnwidth]{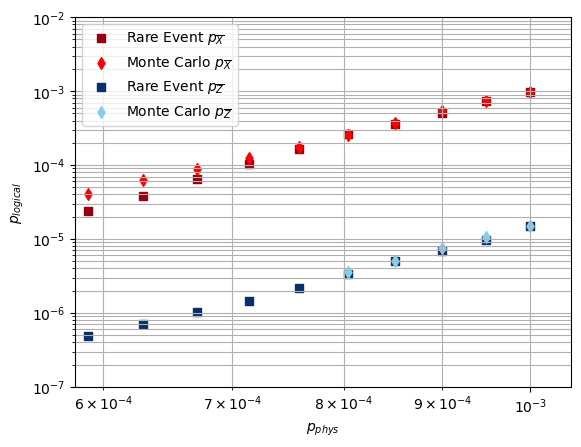}
\caption{Part of the asymmetric noise data in Figure \ref{fig:asymmetric_plot} with $10\times$ the noise in a column and  $\geq100$ jumps  compared to Monte Carlo estimates. The Monte Carlo processes use at least 1,000 failures  to estimate $p_{\overline{X}}$ and  $p_{\overline{Z}}$.} \label{fig:asymmetric_zoom_in}
\end{figure}

For the simulations shown in the previous figures, the circuit noise was symmetrical. However, the rare event simulation easily extends to cases with asymmetrical noise, such as if qubits in certain locations of a surface code are more prone to failure than others, as might be expected in actual hardware (see, e.g., \cite{google2023suppressing}). Figures \ref{fig:asymmetric_plot}(b) and \ref{fig:asymmetric_plot}(c) show logical $X$ ($p_{\overline{X}}$) and $Z$ error rates for a distance-11 rotated surface code in which the checks highlighted in Figure \ref{fig:asymmetric_plot}(a) are more likely to fail. Figure \ref{fig:asymmetric_zoom_in} shows a portion of the same plots with a comparison to estimates found using standard Monte Carlo simulations.

\section{Conclusions and Future Work}\label{sec:future}
\subsection{Future Work}
We identify several directions for future work.

\subsubsection{Confidence and Convergence}
In the rare event simulation, the ratio $C$ satisfying Equation \ref{eqn:c_satisfy} is estimated for each consecutive pair of physical error probabilities in the splitting  sequence. Since the ratios are multiplied to estimate the logical failure rate, errors in these estimates accumulate.  The matters of (1) ensuring that enough jumps have been taken for convergence of the Markov chains and (2) quantifying confidence for estimates remain open. 

A diagnostic for convergence such as the Gelman-Rubin statistic \cite{gelmanRubin} can provide an intuitive, computable test, but does not provide a measure of confidence while also being sensitive to iterative statistics following a long tail or with different within-chain variance.  

\subsubsection{Splitting Heuristics}
To determine sequences of $p$ values, we used the heuristic in Equation \ref{eqn:heuristic} chosen by \cite{bv13} for its tradeoff between the number of splitting steps and statistical error. In cases with asymmetric noise, one could study a tradeoff between statistical error and the number of $p$ values.

 

 \subsubsection{Further Simulations}
 In Sections \ref{sec:leakage} and \ref{sec:postselection}, we described extensions to circuits with leakage and  with post-selection. Implementing these extensions and providing simulated results remains as future work.

\subsection{Conclusions}
We have outlined an approach extending earlier work by Bravyi and Vargo \cite{bv13} to the more complex circuit noise model. For our  extension to the circuit model, we performed simulation for logical failure rates below $10^{-20}$ regime and showed agreement with Monte Carlo simulations in a more accessible ($>10^{-6}$) regime.  Our work has provided a technique for which other simulation software
can be developed for studying teraquop logical failure rates deemed useful for
meaningful quantum computation under realistic noise conditions.

\section{Author Contributions}
All authors contributed substantially to this work. No generative AI models
were used directly in the generation, analysis, or writing of this paper.

\begin{acknowledgments}\label{sec:ack}
We want to express immense gratitude to David Collins and Joseph Maurer at Sandia National Laboratories, who were instrumental in the base
code development that was matured to enable our rare event simulations for circuit-based noise models.  
We
also want to thank Andrew Landahl and Antonio Russo at Sandia National Laboratories for their valuable discussions
and expert support of ideas as they were being developed for this work.

Sandia National Laboratories is a multi-mission laboratory managed and operated by National Technology \& Engineering Solutions of Sandia, LLC (NTESS), a wholly owned subsidiary of Honeywell International Inc., for the U.S. Department of Energy’s National Nuclear Security Administration (DOE/NNSA) under contract DE-NA0003525. This written work is authored by an employee of NTESS. The employee, not NTESS, owns the right, title and interest in and to the written work and is responsible for its contents. Any subjective views or opinions that might be expressed in the written work do not necessarily represent the views of the U.S. Government. The publisher acknowledges that the U.S. Government retains a non-exclusive, paid-up, irrevocable, world-wide license to publish or reproduce the published form of this written work or allow others to do so, for U.S. Government purposes. The DOE will provide public access to results of federally sponsored research in accordance with the DOE Public Access Plan.

This paper describes objective technical results and analysis. Any subjective views or opinions that might be expressed in the paper do not necessarily represent the views of the U.S. Department of Energy or the United States Government.
\end{acknowledgments}

\bibliography{main}

\onecolumn\clearpage
\appendix
\section{\label{sec:appendix}Appendix}
\begin{figure}[!htbp]
\centering
\includegraphics[width=0.9\columnwidth]{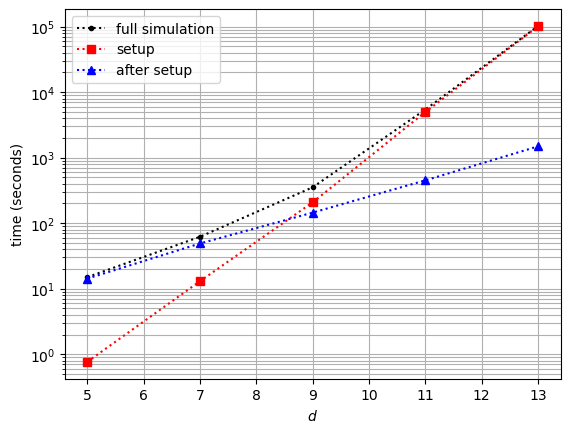}
\caption{Cumulative time to run the simulation with $p_{phys}\in [10^{-4},10^{-3}]$. The setup time includes running a Monte Carlo simulation at $p_{phys}=10^{-3}$ to generate the initial failing events and to get the initial estimates of the logical failure rates, $p_{\overline{X}}$ and $p_{\overline{Z}}$.}\label{fig:mpi-times}
\end{figure}

A breakdown of the runtimes for our implementation of the rare event simulation for odd distance surface codes is
shown in Figure \ref{fig:mpi-times}. As the distance increases, the initial setup, which includes a standard Monte Carlo simulation at the starting point where $p=10^{-3}$ accounts for the bulk of the simulation time.
Note that this starting point is the easiest/fastest for the Monte Carlo simulation because fewer runs are needed to generate
a high quality estimate of the initial logical failure rate for seeding the rare event simulation.
For higher distance codes, starting at a larger $p$ (chosen below the pseudo-threshold\footnote{The pseudo-threshold for a code is the point below which the logical failure rate is below the physical gate failure rate.}) may speed up the simulation. We also note that while the Monte Carlo simulation can be sped up through parallelization (e.g. using MPI or threads), our implementation of the rare event algorithm is not amenable to parallelization within a MCMC chain; however, parallelization helps with the quality of the estimator as used by convergence techniques like the Gelman-Rubin diagnostic~\cite{gelmanRubin}.

\begin{figure}[!ht]\centering
\includegraphics[width=0.75\columnwidth]{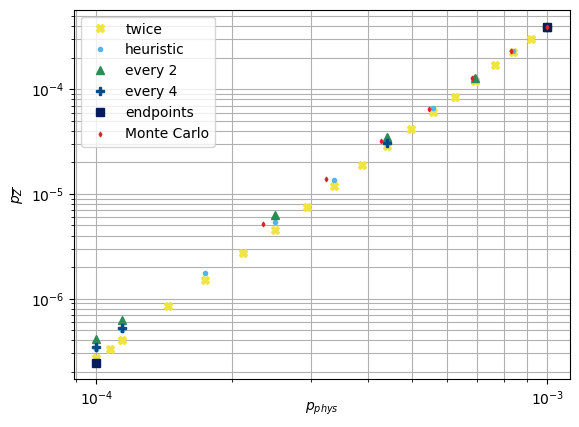}

\caption{Logical $Z$ error rate estimates for a distance-5 rotated surface code for $p_{phys}\in [10^{-4},10^{-3}]$. 
The legend entries refer to the points in the $p_{phys}$ sequence relative to the splitting heuristic. 
}\label{fig:different_pins}
\end{figure}

\begin{figure}[!ht]\centering
\includegraphics[width=0.75\columnwidth]{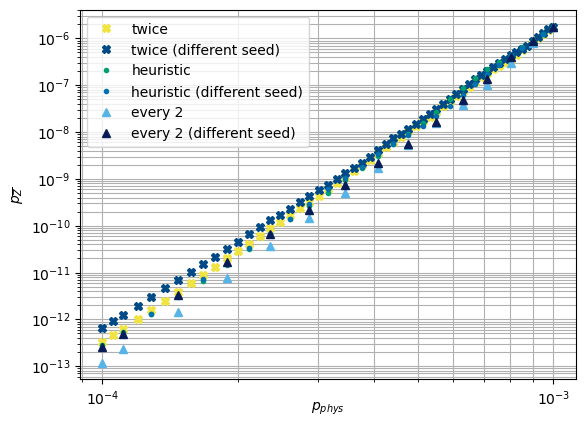}
\caption{Logical $Z$ error rate estimates for a distance-11 rotated surface code for $p_{phys}\in [10^{-4},10^{-3}]$. Estimation using rare event simulation with sequences of physical failure rates. }\label{fig:no_d11_different_pins}
\end{figure}

Figure \ref{fig:different_pins}
shows comparisons of $p_{\overline{Z}}$ estimates using
different sequences of $p$ for the distance-5 rotated surface
code, and Figure \ref{fig:no_d11_different_pins} shows
estimates for the distance-11 rotated surface code. For distance-5,
using different sequences of $p$ yields similar
estimates. For the points labeled  ``twice'' in the plot, a point was added halfway between each point of the heuristic, Equation~\ref{eqn:heuristic}.  For distance-11, there is some difference between
the estimates from different splitting schemes but we suspect
that this is the effect of not yet reaching the steady state
distribution in the MCMC approach.

\end{document}